\documentclass[pra,twocolumn,amsmath,amssymb,10pt,aps,longbibliography,superscriptaddress,citeautoscript,nobibnotes]{revtex4-2}

\usepackage{feynmp}
\usepackage{amsmath}
\usepackage{graphicx}
\usepackage{multirow}
\usepackage{latexsym}
\usepackage{textcomp}
\usepackage{verbatim}
\usepackage{color}
\usepackage{bm}
\usepackage{subfigure}
\usepackage{everysel}
\usepackage{keyval}
\usepackage{ragged2e}
\usepackage{dsfont}
\usepackage{amssymb}
\usepackage{enumitem}
\usepackage{pstricks}
\usepackage{mathtools}

\usepackage{braket}
\usepackage{slashed} 
\usepackage{hyperref}        
\hypersetup{
     colorlinks=true,
     citecolor = red,    
     linkcolor=magenta,
     }

\usepackage{graphicx}
\usepackage{xcolor}
\usepackage{amsmath}
\usepackage{bbold}
\usepackage{amsfonts}
\usepackage{bbm}
\usepackage{comment}
\usepackage{orcidlink}

\newcommand{\osum}{{%
    \setbox0\hbox{\circ}%
    \rlap{\hbox to \wd0{\hss\sum\hss}}\box0
}}

\begin{document}

\title{Universality classes split by strong and weak symmetries}

\author{Jongjun M. Lee\,\orcidlink{0000-0002-9786-1901}}
\thanks{Contact author: jongjun@ualberta.ca}
\affiliation{Department of Physics, University of Alberta, Edmonton, Alberta T6G 2E1, Canada}
\affiliation{Quantum Horizons Alberta \& Theoretical Physics Institute, University of Alberta, Edmonton, Alberta T6G 2E1, Canada}

\author{Myung-Joong Hwang\,\orcidlink{0000-0002-0176-6740}}
\thanks{Contact author: myungjoong.hwang@duke.edu}
\affiliation{Division of Natural and Applied Sciences, Duke Kunshan University, Kunshan, Jiangsu 215300, China}

\author{Igor Boettcher\,\orcidlink{0000-0002-1634-4022}}
\thanks{Contact author: iboettch@ualberta.ca}
\affiliation{Department of Physics, University of Alberta, Edmonton, Alberta T6G 2E1, Canada}
\affiliation{Quantum Horizons Alberta \& Theoretical Physics Institute, University of Alberta, Edmonton, Alberta T6G 2E1, Canada}

\begin{abstract}
Dissipative phase transitions are strongly shaped by the symmetries of the Liouvillian, yet a clear understanding of the quantitative impact of weak and strong symmetries on critical behavior is presently lacking. We study a squeezed-photon model with single- and two-photon losses, realizing weak and strong symmetries in the simplest possible setting. The two symmetries exhibit identical Gaussian static fluctuations, whereas the order parameter and the asymptotic decay rate display distinct scaling behaviors. Our one-loop Keldysh analysis, together with cumulant-expansion numerics, reveals sharply different critical scaling with respect to the thermodynamic scaling parameter. This shows that weak and strong symmetries lead to distinct dynamical universality classes despite originating from the same symmetry group in the closed system. Our results provide a clear quantitative demonstration that strong symmetries fundamentally reshape dissipative criticality.
\end{abstract}

\date{\today}
\maketitle

\section{Introduction}
Recent advances in cavity and circuit quantum electrodynamics have enabled access to ultrastrong and deep-strong light–matter coupling~\cite{gunter2009sub,niemczyk2010circuit,yoshihara2017superconducting,bayer2017terahertz}. At the same time, dissipation can now be engineered with a high degree of control~\cite{poyatos1996quantum,plenio2002entangled,kraus2008preparation,leghtas2015confining}. Together, these developments make driven-dissipative systems a practical platform for studying nonequilibrium collective phenomena. Among them, dissipative phase transitions have attracted considerable attention as they exhibit critical phenomena beyond the equilibrium paradigm~\cite{kessler2012dissipative,carmichael2015breakdown,puebla2017probing,minganti2018spectral,curtis2021critical}. They are marked by diverging time scales and enhanced steady-state fluctuations. Such transitions have been predicted and observed in a range of photonic~\cite{rodriguez2017probing,fink2018signatures,zhang2021driven,beaulieu2025observation}, superconducting circuit~\cite{raftery2014observation,fitzpatrick2017observation,brookes2021critical}, hybrid quantum~\cite{bibak2023dissipative,lee2023cavity,wang2024quantum,lee2025diverging}, trapped ion~\cite{hwang2018dissipative,cai2022probing,sierant2022dissipative}, and cold-atom systems~\cite{baumann2010dicke,brennecke2013real,klinder2015dynamical,lyu2024multicritical}. As dissipative phase transitions become experimentally accessible, a central question arises: which aspects of symmetry can affect universality in dissipative phase transitions, beyond what is expected from closed systems?

In open quantum systems, symmetries can be classified into two distinct types, weak and strong, which have no direct analogue in closed systems~\cite{buvca2012note,albert2014symmetries}. While both types constrain the dynamics and can undergo spontaneous symmetry breaking, they can lead to qualitatively different dissipative behavior. Recent work has highlighted a novel strong-to-weak symmetry breaking and has introduced order parameters capable of diagnosing such transitions~\cite{lee2023quantum,sala2024spontaneous,lessa2025strong,liu2025diagnosing,ziereis2025strong,gu2025spontaneous}. Related studies have further revealed novel topological orders of mixed states~\cite{zhang2025strong,guo2025strong}, emergent hydrodynamic behavior~\cite{huang2025hydrodynamics}, quantum error corrections~\cite{lieu2020symmetry,liu2024dissipative,zhu2024passive}, and measurement schemes~\cite{sun2025scheme,feng2025hardness}. 

Despite this progress, a basic issue remains unresolved. It is still unclear whether weak and strong symmetries, which are closely related at the level of the symmetry group, lead to genuinely different critical behavior in dissipative phase transitions. In particular, the extent to which the nature of symmetry controls universality in open quantum criticality has not yet been systematically established~\cite{sieberer2013dynamical,altland2021symmetry,sa2023symmetry,sieberer2025universality}.

\begin{figure}[t!]
    \centering
    \includegraphics[width=1.0\linewidth]{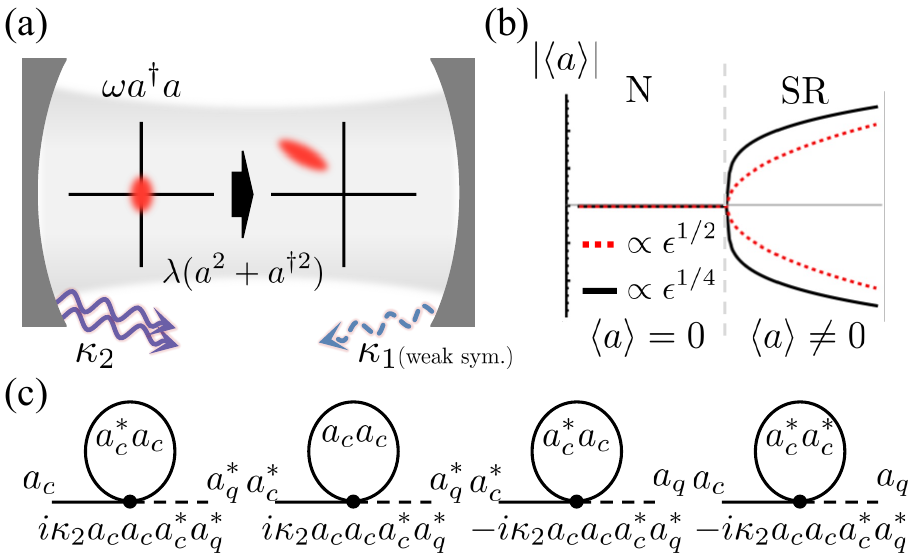}
    \caption{Schematic illustration of (a) the system and (c) the loop diagrams relevant to the quartic-order corrections to the retarded Green function. Panel (b) shows the phase diagram as a function of the two-photon driving strength $\lambda$, obtained from the numerical data~\cite{appx_ref}. In panel (a), the single-photon loss ($\kappa_{1}$) is present only for the weak parity symmetry. The plot inside the cavity schematically shows the Wigner function in phase space and illustrates symmetry breaking. In panel (b), the order parameter $\langle a\rangle$ in the superradiant phase follows the scaling with the exponent given in Eq.~\eqref{Eq_a_1}, where $\epsilon=|\lambda-\lambda_{c}|$. N and SR denote the normal and superradiant phases, respectively.}
    \label{fig1}
\end{figure}

In this work, we focus on strong and weak parity symmetries in open quantum systems. These two symmetries share a common origin in the parity symmetry of the underlying closed system~\cite{hwang2015quantum,kirton2019introduction}, but differ in how the symmetry is implemented once dissipation is present~\cite{lieu2020symmetry}. We study a minimal model of a single-mode cavity photon subject to squeezing, with single- and two-photon losses providing the dissipative channels~\cite{mirrahimi2014dynamically,bartolo2016exact,minganti2016exact,roberts2020driven,lieu2020symmetry,mylnikov2022dissipative,mylinikov2025emergent}; see Fig.~\ref{fig1}(a). By tuning the single-photon loss, we selectively realize weak or strong parity symmetry, while the squeezing strength controls the dissipative phase transition.

We show that weak and strong parity symmetries give rise to distinct universality classes in dissipative phase transitions. Although the two symmetries exhibit identical Gaussian fluctuations near criticality, their scaling behavior with the thermodynamic scaling parameter differs sharply beyond the Gaussian level~\cite{dalla2012dynamics,mathey2019absence,kang2025nongaussian}. Using a Keldysh path-integral analysis and an equivalent Langevin description, we obtain closed-form expressions for the critical behavior~\cite{torre2013keldysh,sieberer2016keldysh,mari2012cooling,woolley2014two}. These analytical results are further supported by numerics based on a cumulant expansion of the equations of motion derived from the master equation~\cite{kubo1962generalized,reiter2020cooperative,fowler2023determining}.

\section{Model and symmetry}
A single photonic mode is subject to two-photon driving and coupled to an environment via single- and two-photon losses. The dynamics of the density matrix $\rho$ is governed by the Lindblad master equation,
\begin{equation}
\frac{\partial \rho}{\partial t} = \mathcal{L}[\rho] = -i[H,\rho]+ \sum_{j=1,2} \mathcal{D}[L_{j}],
\label{Eq_Master_1}
\end{equation}
with the Hamiltonian
\begin{equation}
    H = \omega a^{\dagger}a + \lambda(a^{2}+a^{\dagger 2}).
\end{equation}
The dissipators are defined as $L_{1}=\sqrt{\kappa_{1}}a$ and $L_{2}=\sqrt{\kappa_{2}}a^{2}$, and $\mathcal{D}[L]=2L\rho L^{\dagger}-L^{\dagger}L\rho - \rho L^{\dagger}L$. Here, $\omega>0$ is the photon frequency, $\lambda\ge 0$ denotes the two-photon driving strength, and $\kappa_{1}\ge 0$ and $\kappa_{2}>0$ are the single- and two-photon loss rates, respectively. A schematic of the system is shown in Fig.~\ref{fig1}(a). We will use the notation $\langle \mathcal{O}\rangle = \text{Tr}[\mathcal{O}\rho_{\rm ss}]$ where $\rho_{\rm ss}$ is the steady state, i.e., $\mathcal{L}[\rho_{\rm ss}]=0$. 

The Lindbladian exhibits a parity symmetry whose nature depends on the presence of the single-photon loss $\kappa_{1}$. We define the parity operator $\Pi=\exp(i\pi a^{\dagger}a)$, which satisfies $\{\Pi,L_{1}\}=0$ and $[\Pi,H]=[\Pi,L_{2}]=0$~\cite{albert2014symmetries}. We also define the left and right superoperators $\Pi_{l}=\Pi(\cdot)$ and $\Pi_{r}=(\cdot)\Pi^{\dagger}$, which act on the ket and bra, respectively~\cite{lieu2020symmetry}. As a consequence, with finite single-photon loss ($\kappa_{1}\neq 0$), the Lindbladian exhibits a weak parity symmetry, $\mathcal{L}[\Pi \rho \Pi^{\dagger}] = \Pi \mathcal{L}[\rho]\Pi^{\dagger}$, which follows from (anti-) commutation relations above. By contrast, in the absence of single-photon loss ($\kappa_{1}=0$), it possesses a strong parity symmetry, $\mathcal{L}[\Pi_{l(r)}[\rho]]=\Pi_{l(r)}[\mathcal{L}[\rho]]$, since $\Pi$ commutes with both Hamiltonian and the dissipator~\cite{lieu2020symmetry}. In the presence of these symmetries, the Lindbladian decomposes into symmetry-resolved sectors, with the strong parity symmetry giving rise to decoherence-free subspaces that can protect quantum information against dissipation~\cite{zanardi1997noiseless,lidar1998decoherence}. Accordingly, in this open quantum system, the presence or absence of single-photon loss leads to a realization of either a strong or a weak parity symmetry.

\section{Order parameter}
This model undergoes a dissipative phase transition at the critical driving strength $\lambda_c=\frac{1}{2}\sqrt{\omega^{2}+\kappa_{1}^{2}}$~\cite{lieu2020symmetry}. For $\lambda<\lambda_c$, the parity symmetry is preserved and no photon condensation occurs in the steady state, $\langle a\rangle=0$, corresponding to the normal phase. For $\lambda>\lambda_c$, the parity symmetry is spontaneously broken, and a macroscopic photon condensate develops as an order parameter in the steady state~\cite{mylnikov2022dissipative}. Near the critical point, the order-parameter amplitude obtained from mean-field theory is given by
\begin{equation}
|\langle a \rangle| \simeq
\begin{cases}
\sqrt{\frac{\lambda_{c}}{\kappa_{1}\kappa_{2}}}\sqrt{\lambda-\lambda_{c}}, & (\kappa_{1}\neq 0), \\
\frac{\omega^{1/4}}{\sqrt{2\kappa_{2}}}(\lambda-\lambda_{c})^{1/4}, & (\kappa_{1}=0).
\end{cases}
\label{Eq_a_1}
\end{equation}
corresponding to weak and strong parity symmetries, respectively. Here, mean-field theory refers to the approximation in which quantum correlations are neglected in the thermodynamic limit. This phase is referred to as the superradiant phase. We schematically illustrate the corresponding phase diagram in Fig.~\ref{fig1}(b). 

For the dissipative phase transition considered here, the two-photon loss rate $\kappa_{2}$ plays the role of an effective thermodynamic scaling parameter. A finite $\kappa_{2}$ prevents true photon condensation by inducing a finite-size crossover~\cite{carmichael2015breakdown,curtis2021critical}, playing a role analogous to that of the inverse system size $1/N$ in the Dicke model~\cite{kirton2019introduction}. A sharp phase transition emerges only in the limit of vanishing $\kappa_{2}$~\cite{zhu2024passive,jayesh2025dissipative}.

\section{Fluctuation and critical scaling}
To investigate Gaussian fluctuations near the transition, we employ the Keldysh path-integral formalism. Gaussian fluctuations refer to fluctuations described up to quadratic order around the mean-field solution~\cite{serafini2023quantum}. Starting from the Lindblad master equation in Eq.~(\ref{Eq_Master_1}), the quadratic (Gaussian) Keldysh action takes the form
\begin{equation}
    S^{(2)} = \frac{1}{2}\int \frac{d\nu}{2\pi} \Psi^{\dagger}_{\nu}
    \begin{pmatrix}
        0 & [G^{A}_{0}(\nu)]^{-1} \\[2pt]
        [G^{R}_{0}(\nu)]^{-1} & \Sigma^{K}_{0}
    \end{pmatrix}
    \Psi_{\nu},
\label{Eq_S2_1}
\end{equation}
where $\Psi^{\dagger}_{\nu}= (\delta a_{c,\nu},\delta a^{*}_{c,-\nu},\delta a_{q,\nu},\delta a^{*}_{q,-\nu} )^{\rm T}$ is the Nambu-Keldysh spinor~\cite{sieberer2016keldysh}. Here, $\delta a = a-\langle a\rangle$ denotes fluctuations around the mean-field solution, and the classical and quantum components are defined as $a_{c/q}=(a_{+}\pm a_{-})/\sqrt{2}$. The retarded Green function is given by
\begin{equation}
    [G^{R}_{0}(\nu)]^{-1} =
    \begin{pmatrix}
        \nu-\omega+i\Gamma & -2\lambda+2i\rho\\
        -2\lambda-2i\rho & -\nu-\omega-i\Gamma
    \end{pmatrix},
\label{Eq_G_R_1}
\end{equation}
with the advanced Green function $G^{A}_{0}(\nu)=G^{R}_{0}(\nu)^{\dagger}$. The Keldysh self-energy is $\Sigma^{K}_{0}=2i\Gamma \mathcal{I}_{2}$, where $\mathcal{I}_{2}$ is the $2\times2$ identity matrix. The effective damping rate is $\Gamma=\kappa_{1}+4\rho$, with $\rho=\kappa_{2}|\langle a\rangle|^{2}$.

The number and squeezing fluctuations, $\delta n=\langle\delta a^{\dagger}\delta a\rangle$ and $\delta m=\langle\delta a^{2}\rangle$, are obtained from the Keldysh Green function $G^{K}_{0}(\nu)=G^{R}_{0}(\nu)\Sigma^{K}_{0}G^{A}_{0}(\nu)$ as
\begin{equation}
\begin{aligned}
    \delta n =& -\frac{i}{2}\int \frac{d\nu}{2\pi}[G^{K}_{0}(\nu)]_{22}-\frac{1}{2},\\
    \delta m =& -i \int \frac{d\nu}{2\pi} [G^{K}_{0}(\nu)]_{12}.
\end{aligned}
\label{Eq_nm_1}
\end{equation}
Near the critical point $\lambda=\lambda_{c}$, Gaussian fluctuations diverge algebraically and exhibit power-law behavior. For the weak parity symmetry, we find
\begin{equation}
    \delta n,\; \mathrm{Re} [\delta m ],\; \mathrm{Im} [\delta m ] \propto \frac{1}{|\lambda-\lambda_{c}|}.
\end{equation}
For the strong parity symmetry, we find
\begin{equation}
    \delta n,\; \mathrm{Re} [\delta m ] \propto \frac{1}{|\lambda-\lambda_{c}|},
\end{equation}
whereas
\begin{equation}
\mathrm{Im} [\delta m ] \propto
\begin{cases}
0, & (\lambda <\lambda_{c}), \\
\frac{1}{\sqrt{\lambda-\lambda_{c}}}, & (\lambda >\lambda_{c}).
\end{cases}
\end{equation}
Thus, most quantities show a power-law divergence with exponent $-1$, while $\mathrm{Im}[\delta m]$ for the strong parity symmetry vanishes in the normal phase at the Gaussian level due to the structure of the quadratic action and instead exhibits a weaker divergence with exponent $-1/2$ in the superradiant phase. The purity $\mu\equiv \text{Tr}[\rho^{2}]$ of the steady state shows a square-root scaling on both sides of the transition, i.e., $\mu\propto\sqrt{|\lambda-\lambda_{c}|}$. The exact coefficients of these divergences are summarized in Table~\ref{Table_summary_1}.

\begin{table}[t!]
\begin{tabular}{cccccc}
\hline \hline 
& $\delta n$   & $\text{Re}[\delta m]$ & $\text{Im}[\delta m]$ & $\mu$ & $\Gamma_{\rm ADR}$\\ \hline
W, N & $\frac{\lambda_{c}}{4}\epsilon^{-1}$ & $-\frac{\omega}{8}\epsilon^{-1}$ & $-\frac{\kappa_{1}}{8}\epsilon^{-1}$ & $\sqrt{\frac{2}{\lambda_{c}}}\epsilon^{1/2}$ & $\frac{4\lambda_{c}}{\kappa_{1}}\epsilon$ \\ \hline
W, SR & $\frac{\lambda_{c}}{4}\epsilon^{-1}$ & $-\frac{\omega}{8}\epsilon^{-1}$ & $-\frac{\kappa_{1}}{8}\epsilon^{-1}$ & $\sqrt{\frac{2}{\lambda_{c}}}\epsilon^{1/2}$ & $\frac{4\lambda_{c}}{\kappa_{1}}\epsilon$ \\ \hline
S, N & $\frac{\omega}{8}\epsilon^{-1}$ & $-\frac{\omega}{8}\epsilon^{-1}$ & $0$ & $\frac{2}{\sqrt{\omega}}\epsilon^{1/2}$ & $0$ \\ \hline
S, SR & $\frac{\omega}{16}\epsilon^{-1}$ & $-\frac{\omega}{16}\epsilon^{-1}$ & $-\frac{\sqrt{\omega}}{8}\epsilon^{-1/2}$ & $\sqrt{\frac{8}{\omega}}\epsilon^{1/2}$ & $4\sqrt{\omega}\epsilon^{1/2}$ \\ \hline
\end{tabular}
\caption{Summary of the scaling behavior of Gaussian fluctuations as a function of the two-photon driving strength $\lambda$ near the critical point $\lambda_c$. Here, $\epsilon\equiv |\lambda-\lambda_{c}|$. W and S label the weak and strong parity symmetries, respectively, while N and SR denote the normal and superradiant phases.}
\label{Table_summary_1}
\end{table}

At the Gaussian level, both weak and strong symmetries therefore exhibit identical critical exponents for the divergence of number and squeezing fluctuations, as well as for the vanishing of the purity, except for $\text{Im}[\delta m]$~\cite{sieberer2016keldysh}. Experimentally, measured quadrature variances involve a linear combination of $\mathrm{Re}[\delta m]$ and $\mathrm{Im}[\delta m]$ and may therefore follow the scaling behavior of the former contribution, with exponent $-1$, in most cases~\cite{lvovsky2009continuous}. Isolating the scaling of $\mathrm{Im}[\delta m]$ would probe the distinctive critical behavior of the strong symmetry already at the Gaussian level; devising such a measurement scheme is therefore an interesting direction for future work. Consequently, the two symmetry classes differ only in nonuniversal coefficients and in specific correlators, rather than in the universal critical scaling. This demonstrates that Gaussian static fluctuations are insensitive to the distinction between strong and weak symmetries here, implying that symmetry-dependent critical behavior of static fluctuations must originate beyond the quadratic approximation.

\section{Asymptotic decay rate}
The asymptotic decay rate $\Gamma_{\rm ADR}$ (ADR), often identified with the smallest nonzero decay rate in the Liouvillian spectrum~\cite{buvca2012note}, controls the slowest relaxation toward the steady state. At the Gaussian level, it is determined by the poles of the retarded Green function in Eq.~(\ref{Eq_G_R_1}). For the weak parity symmetry $(\kappa_{1}\neq 0)$, the ADR near the critical point exhibits linear scaling on both sides of the transition and is given by
\begin{equation}
\Gamma_{\rm ADR}(\kappa_{1}\neq 0) \simeq \frac{4\lambda_{c}}{\kappa_{1}}|\lambda-\lambda_{c}|.
\end{equation}

In contrast, the behavior of the ADR for the strong parity symmetry $(\kappa_{1}=0)$ is qualitatively different. In the normal phase, the retarded Green function $G^{R}_{0}(\nu)$ becomes purely real, implying the absence of dissipative relaxation channels~\cite{buvca2012note,albert2014symmetries}. As a consequence, all decay rates vanish in the thermodynamic limit, and the ADR is not well defined. Physically, this reflects the emergence of an extensive manifold of stationary states with infinite lifetime~\cite{sanchez2019symmetries,joseph2023on}. By contrast, in the superradiant phase, the ADR becomes finite and follows the scaling
\begin{equation}
\Gamma_{\rm ADR}(\kappa_{1}=0) \simeq 4\sqrt{\omega}\sqrt{\lambda-\lambda_{c}}.
\end{equation}
This square-root behavior sharply contrasts with the linear scaling found for weak parity symmetry. While this difference signals that systems with weak and strong parity symmetries belong to different universality classes, a systematic analysis of non-Gaussian fluctuations is required to further investigate the corresponding universality classes. The above results are summarized in Table~\ref{Table_summary_1}.

\section{Quartic-order correction}
We now investigate the non-Gaussian fluctuations arising from higher-order corrections at the critical point. The higher-order corrections refer to terms beyond quadratic order in the fluctuations. Since there is no condensation at this point, the quartic-order action originating from the two-photon loss can be written directly in terms of the bare Keldysh fields $a_{c/q}$ as
\begin{equation}
    S^{(4)} = i \int dt \left(
    L^{\dagger}_{q}L_{c}
    - L^{\dagger}_{c}L_{q}
    + 2L^{\dagger}_{q}L_{q}
    \right),
\end{equation}
with $L_{c}= \sqrt{\kappa_{2}/2} (a^{2}_{c}+a^{2}_{q})$ and $L_{q}=\sqrt{2\kappa_{2}} a_{c} a_{q}$. This quartic action generates five distinct types of Keldysh vertices. We treat these interaction effects perturbatively, assuming that $\kappa_{2}$ is small.

Among these vertices, those involving three classical fields and one quantum field yield corrections to the quadratic Keldysh action at the one-loop level; see Fig.~\ref{fig1}(c). As a representative example, the retarded Green function acquires the form
\begin{equation}
    [G^{R}(\nu)]^{-1} =
    \begin{pmatrix}
        \nu-\omega+i\tilde{\Gamma} & -2\lambda+2i\tilde{\rho}^{*}\\
        -2\lambda-2i \tilde{\rho} & -\nu-\omega-i\tilde{\Gamma}
    \end{pmatrix},
\label{Eq_GR_Corr_1}
\end{equation}
where $\tilde{\Gamma}=\Gamma+2\kappa_{2}n$ and $\tilde{\rho} = \rho+\kappa_{2}m/2$. Using this one-loop-corrected quadratic action, we compute both the static and dynamical fluctuations and obtain analytic expressions at the critical point as functions of $\kappa_{2}$. The result for the strong parity symmetry case is consistent with Ref.~\cite{mylinikov2025emergent}, based on the $P$ representation and Boltzmann-Gibbs theory. The exact coefficients of these results are summarized in Table~\ref{Table_summary_2}.

These results reveal that the non-Gaussian fluctuations exhibit distinct scaling behaviors with respect to $\kappa_{2}$, which plays the role of an effective thermodynamic scaling parameter, depending on whether the system possesses weak or strong parity symmetry. For instance, with the weak parity symmetry, the number fluctuation scales as $1/\sqrt{\kappa_{2}}$, whereas with the strong parity symmetry, it scales as $1/\kappa_{2}^{2/3}$. This implies that the two symmetries belong to different universality classes, despite sharing a common symmetry operator in the corresponding closed system. This distinction constitutes the central result of this work.

We note that, in the strong-symmetry case, the exact Liouvillian gap of the full problem with finite $\kappa_{2}$ is always zero because the Liouvillian decomposes into invariant symmetry sectors with independent stationary states~\cite{buvca2012note,albert2014symmetries}. Therefore, in the strong-coupling regime, the scaling of the ADR with respect to $\kappa_{2}$ given in Table~\ref{Table_summary_2} is not the exact Liouvillian gap of the full system with finite $\kappa_{2}$, but the nonzero smallest decay rate associated with the symmetry-broken branch in the thermodynamic-limit description.

\begin{table}[t!]
\begin{tabular}{cccccc}
\hline \hline 
& $\delta n$   & $\text{Re}[\delta m]$ & $\text{Im}[\delta m]$ & $\mu$ & $\Gamma_{\rm ADR}$\\ \hline
W & $\frac{\lambda_{c}}{\sqrt{\kappa_{1}\kappa_{2}}}$ & $-\frac{\omega}{2\sqrt{\kappa_{1}\kappa_{2}}}$ & $-\frac{1}{2}\sqrt{\frac{\kappa_{1}}{\kappa_{2}}}$ & $\frac{1}{2\sqrt{\lambda_{c}}} (\frac{\kappa_{2}}{\kappa_{1}} )^{1/4}$ & $2\lambda_{c}\sqrt{\frac{\kappa_{2}}{\kappa_{1}}}$ \\ \hline
S & $\frac{\omega^{2/3}}{\sqrt{6}\kappa^{2/3}_{2}}$ & $-\frac{\omega^{2/3}}{\sqrt{6}\kappa^{2/3}_{2}}$ & $-\frac{\omega^{1/3}}{2\kappa^{1/3}_{2}}$ & $\sqrt{\frac{3}{2\sqrt{6}-3}}(\frac{\kappa_{2}}{\omega})^{1/3}$ & $\frac{2\omega^{2/3}}{\sqrt{6}}\kappa^{1/3}_{2}$ \\ \hline
\end{tabular}
\caption{Summary of the scaling behavior of non-Gaussian fluctuations as a function of the two-photon loss strength $\kappa_{2}$ at the critical point $\lambda=\lambda_{c}$. W and S label the weak and strong parity symmetries, respectively.}
\label{Table_summary_2}
\end{table}

\section{Universality classes}
We now have analytic expressions for the scaling behaviors, both in the vicinity of the critical point as a function of the two-photon driving strength $\lambda$ and at the critical point as a function of $\kappa_{2}$; they are summarized in Tables~\ref{Table_summary_1} and~\ref{Table_summary_2}. Together with the scaling hypothesis, these results imply the universal scaling form of the photon number fluctuation~\cite{fisher1972scaling,botet1982size}
\begin{equation}
    \delta n (\lambda,\kappa_{2}) = \kappa^{-\zeta_{x}}_{2}F_{\rm w(s)} \Big(|\lambda-\lambda_{c}|^{\nu_{x}}\kappa^{-\zeta_{x}}_{2} \Big)
\label{Eq_Uni_Scal_1}
\end{equation}
where the photon-flux exponent is $\nu_{x}=1$ for both symmetries~\cite{torre2013keldysh}, while $\zeta_{x}=1/2$ and $2/3$ for weak and strong parity symmetries, respectively. Here, $F_{\rm w(s)}(x)$ is a dimensionless scaling function. The Gaussian critical behavior is recovered in the limit $\kappa_{2}\to 0$ at fixed $\lambda\neq\lambda_{c}$, which requires $F_{\rm w(s)}(x\to\infty)\propto x^{-1}$. At the critical point $\lambda=\lambda_{c}$ with fixed $\kappa_{2}>0$, the scaling function approaches a constant, consistently reproducing the non-Gaussian scaling behavior.

This scaling form already indicates that the systems with weak and strong parity symmetries belong to distinct universality classes. While the individual critical exponents $\nu_{x}$ and $\zeta_{x}$ depend on the observable, their ratio $\xi_{x}=\nu_{x}/\zeta_{x}$ defines the coherence number, which has been argued to be observable-independent in several zero-dimensional driven-dissipative models~\cite{botet1982size,hwang2018dissipative,bettmann2025thermodynamic}. We obtain $\xi_{x}=2$ for weak parity symmetry and $\xi_{x}=3/2$ for strong parity symmetry. Dynamical scaling further supports this classification. The ADR yields the dynamical exponent $\nu_{t}=1,(1/2)$ with respect to $\lambda$ and the cut-off dynamical exponent $\zeta_{t}=1/2,(1/3)$ with respect to $\kappa_{2}$ for weak (strong) parity symmetry~\cite{torre2013keldysh,hohenberg1997theory}. The ratio $\nu_{t}/\zeta_{t}$ again gives $2$ and $3/2$ for the weak and strong parity symmetries, in agreement with the coherence numbers extracted from the number fluctuations. From Tables.~\ref{Table_summary_1} and \ref{Table_summary_2}, one can verify that the ratio of critical exponents is likewise $2$ and $3/2$ for all other quantities listed there, for weak and strong parity symmetries, respectively, except for $\mathrm{Im}[\delta m]$ in the normal phase with the strong symmetry.

Taken together, these results show that the distinction between the universality classes in our model is governed by whether the parity symmetry is weak or strong. The system with weak parity symmetry may belong to the same universality class as the open Dicke and Rabi models with single-photon loss~\cite{torre2013keldysh,hwang2018dissipative}, whereas the model with strong parity symmetry defines a distinct universality class. A more detailed characterization of the universality class, especially in the presence of strong symmetry, would be an interesting direction for future work.

\section{Numerical verification}
Starting from the Lindblad master equation in Eq.~(\ref{Eq_Master_1}), we derive the equations of motion for the first and second moments. We have
\begin{equation}
\begin{aligned}
\frac{d\langle a\rangle}{dt} =& -(i\omega+\kappa_{1})\langle a\rangle -2i\lambda \langle a\rangle^{*} -2\kappa_{2}\langle a^{\dagger}a^{2}\rangle,\\
\frac{d m}{dt}=& -2(i\omega+\kappa_{1}+\kappa_{2})m -2i\lambda -4i\lambda n\\
&-4\kappa_{2}\langle a^{\dagger}a^{3} \rangle,\\
\frac{d n}{dt} =& +2i\lambda(m - m^{*}) -2\kappa_{1} n -4\kappa_{2}\langle a^{\dagger 2}a^{2} \rangle,
\end{aligned}
\label{Eq_EOM_1}
\end{equation}
where $n=\langle a^{\dagger}a\rangle$ and $m=\langle a^{2}\rangle$. In the steady state, the left-hand sides vanish. We truncate the hierarchy using a cumulant expansion up to second order~\cite{fowler2023determining,reiter2020cooperative}, which yields three coupled nonlinear equations for $\langle a\rangle$, $n$, and $m$. See the Appendix for the detailed derivation~\cite{appx_ref}. We note that these equations explicitly depend on $\kappa_{2}$, reflecting the inclusion of non-Gaussian effects arising from the loop corrections shown in Fig.~\ref{fig1}(c), beyond the quadratic approximation.

\begin{figure}[t!]
    \centering
    \includegraphics[width=1.0\linewidth]{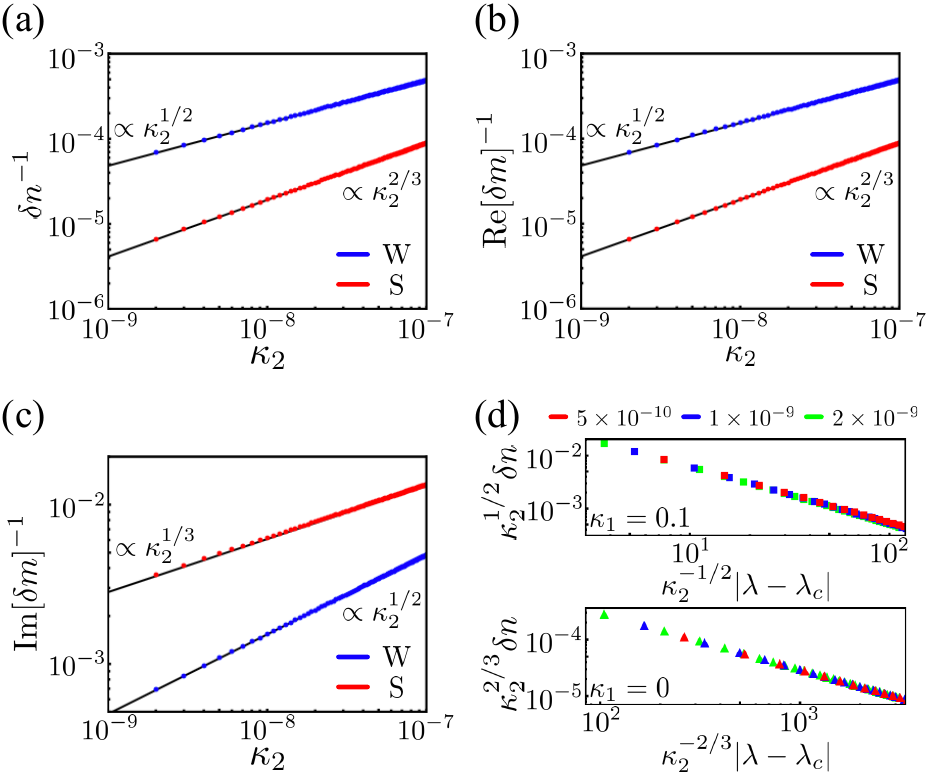}
    \caption{Numerical results from the cumulant expansion. (a) Number fluctuation $\delta n$, (b) $\mathrm{Re}[\delta m]$, and (c) $\mathrm{Im}[\delta m]$ at the critical driving strength $\lambda_c$ as functions of the two-photon loss rate $\kappa_2$ for weak (W) and strong (S) parity symmetries. Black lines indicate power-law fits. (d) Rescaled number fluctuation $\kappa_2^{1/2(2/3)} \delta n$ versus $\kappa_2^{-1/2(-2/3)} |\lambda-\lambda_c|$ in the superradiant phase for weak (strong) parity symmetries. Colors denote different values of $\kappa_2$. The collapse of data points onto a single curve confirms Eq.~(\ref{Eq_Uni_Scal_1}). For the weak-parity-symmetry plots, we set $\kappa_{1}=0.1$. All frequency scales are measured in units of $\omega=1$.} 
    \label{fig2}
\end{figure}

We solve the resulting equations numerically. The solutions reproduce the divergence with respect to the two-photon driving strength $\lambda$, following the power-law behavior summarized in Table~\ref{Table_summary_1}. See the Appendix for the corresponding data~\cite{appx_ref}. In Figs.~\ref{fig2}(a)-(c), we plot the fluctuations at the critical point $\lambda=\lambda_c$ as functions of $\kappa_{2}$. Here, the inverse two-photon loss rate plays the role of an effective thermodynamic scaling parameter. The log-log plots clearly confirm the critical exponents analytically predicted in Table~\ref{Table_summary_2} for both weak and strong parity symmetries.

In Fig.~\ref{fig2}(d), we present the rescaled number fluctuation $\kappa_{2}^{1/2(2/3)} \delta n$ as a function of $\kappa_{2}^{-1/2(2/3)} |\lambda-\lambda_c|$ for the weak(strong) parity symmetry. Note that data obtained for different values of $\kappa_{2}$ collapse onto a single curve. This provides numerical evidence for a universal scaling behavior near the critical point [Eq.~(\ref{Eq_Uni_Scal_1})].

\section{Experimental platforms}
The present model is directly relevant to superconducting bosonic platforms based on two-photon-driven or two-photon-dissipative Kerr resonators~\cite{hajr2024high,beaulieu2025observation}. In particular, dissipative phase transitions have already been observed in a two-photon-driven superconducting Kerr resonator~\cite{beaulieu2025observation}. The two-photon driving can be induced by a parametric pump~\cite{clerk2010introduction}, while the two-photon loss can be realized by coupling the cavity mode to an auxiliary nonlinear dissipative channel~\cite{leghtas2015confining,berdou2023one}, with residual single-photon loss arising from intrinsic decay~\cite{blais2021circuit}. 

Importantly, recent circuit-QED experiments have already reached a regime in which the engineered two-photon loss rate exceeds the residual single-photon loss rate by about two orders of magnitude, namely $\kappa_{2}/\kappa_{1}\gtrsim 10^{2}$~\cite{marquet2024autoparametric}. In that experiment, three-wave mixing was used to realize an autoparametric two-photon dissipative process with $\kappa_{2}/2\pi \simeq 2\,{\rm MHz}$, leading to a stabilized cat state with a long bit-flip time~\cite{marquet2024autoparametric}. Since the residual single-photon loss can, in principle, be further suppressed by Purcell-filter engineering~\cite{jeffrey2014fast,hajr2024high}, these platforms provide a promising setting for experimentally distinguishing the critical phenomena associated with weak and strong parity symmetries.

\section{Conclusion and outlook}
In conclusion, we have demonstrated that weak and strong parity symmetries exhibit distinct critical scaling behaviors and belong to different universality classes, despite sharing a common origin in the parity symmetry of the underlying closed system. Our results show that universality classes in dissipative phase transitions cannot be inferred solely from the symmetry group of the closed system, but instead depend crucially on how that symmetry is implemented at the level of the Lindbladian dynamics, i.e., through weak and strong symmetry realizations.

The distinction between weak and strong symmetry implementations identified here provides a concrete criterion for organizing dissipative universality classes. A pressing future direction is to examine whether this criterion extends to other symmetry settings and to spatially extended driven-dissipative systems. These predictions can be tested experimentally in cavity and circuit QED platforms where dissipative channels are controllable.

While the present work focuses on the critical phenomena of steady states, driven-dissipative systems can, in general, also exhibit nonstationary long-time behavior. Extending the present analysis to the criticality of such oscillatory long-time states would further broaden the scope of this problem~\cite{kopylov2015dissipative,iemini2018boundary,kessler2019emergent}. In particular, it would be interesting to explore situations in which extended symmetries enforce such nonstationary states~\cite{buvca2019non}.

\acknowledgements
We thank Min Ju Park and Canon Sun for inspiring discussions. J.M.L. and I.B. gratefully acknowledge support from Quantum Horizons Alberta. M.-J.H. was supported by the Innovation Program for Quantum Science and Technology 2021ZD0301602. I.B. acknowledges funding from the Natural Sciences and Engineering Research Council of Canada (NSERC) Discovery Grants RGPIN-2021-02534 and DGECR2021-00043.

\appendix

\section*{APPENDIX A: Langevin equation approach}
As an alternative to the Keldysh calculation of the Gaussian fluctuations, we analyze here the Gaussian quantum fluctuations around the steady state by linearizing the Langevin equations and solving the corresponding Lyapunov equation for the steady-state covariance matrix.

We consider fluctuations $\delta a = a-\langle a\rangle$ around the mean-field solution and introduce the quadrature fluctuation operators
\begin{equation}
X=\frac{\delta a+\delta a^{\dagger}}{\sqrt{2}},\qquad
P=\frac{\delta a-\delta a^{\dagger}}{i\sqrt{2}}.
\end{equation}
Starting from the Lindblad master equation [Eq.~(\ref{Eq_Master_1})], the linearized Langevin equation for the quadrature fluctuations takes the form
\begin{equation}
\dot{\mathbf{R}} = A\mathbf{R} + \bm{\eta}, \:
\langle \bm{\eta}(t)\bm{\eta}^{\mathrm T}(t')\rangle = D\delta(t-t'),
\end{equation}
where $\mathbf{R}=(X,P)^{\mathrm T}$. The drift matrix $A$ is obtained by linearizing both the coherent Hamiltonian dynamics
and the nonlinear dissipative terms. We have
\begin{equation}
A=
\begin{pmatrix}
6\rho & \omega-2\lambda \\
-\omega-2\lambda & 2\rho
\end{pmatrix},
\end{equation}
with $\rho=\kappa_{2}|\langle a\rangle|^{2}$. In the superradiant phase, the two-photon loss generates quadrature-dependent effective damping once the condensate amplitude $\langle a\rangle$ becomes finite. The noise vector $\bm{\eta}=(\eta_{X},\eta_{P})^{\mathrm T}$ is defined as
\begin{equation}
\eta_{X}=\frac{a_{\rm in}+a_{\rm in}^{\dagger}}{\sqrt{2}},\:
\eta_{P}=\frac{a_{\rm in}-a_{\rm in}^{\dagger}}{i\sqrt{2}},
\end{equation}
where
\begin{equation}
a_{\rm in}=\sqrt{\kappa_{1}}a_{1,{\rm in}}+2\sqrt{\rho}a_{2,{\rm in}},
\end{equation}
and the input noise operators satisfy
$\langle a_{i,{\rm in}}(t)a^{\dagger}_{j,{\rm in}}(t')\rangle=\delta_{ij}\delta(t-t')$. The noise correlations follow from the linearized dissipator and correspond to vacuum input noise associated with single- and two-photon loss channels. The corresponding diffusion matrix is isotropic, $D=\Gamma \mathcal{I}_{2}$, where $\Gamma=\kappa_{1}+4\rho$. This Langevin description fully captures the Gaussian fluctuations around the mean-field solution.

We now compute the static Gaussian fluctuations. The covariance matrix of the quadratures is defined as
\begin{equation}
V=
\begin{pmatrix}
\langle X^{2}\rangle & \frac{1}{2}\langle XP+PX\rangle \\
\frac{1}{2}\langle XP+PX\rangle & \langle P^{2}\rangle
\end{pmatrix}.
\end{equation}
For a Gaussian bosonic system, $V$ satisfies the Lyapunov equation~\cite{mari2012cooling,woolley2014two}.
\begin{equation}
AV+VA^{\mathrm T}+D=0.
\label{Eq_Lyapunov_1}
\end{equation}
Solving Eq.~(\ref{Eq_Lyapunov_1}), we obtain closed forms
\begin{equation}
\begin{aligned}
\langle X^{2}\rangle &=
\frac{c_{3}(\Omega_{1}^{2}+\Omega_{1}\Omega_{2}+c_{1}c_{2}+c_{2}^{2})}
{2(c_{1}+c_{2})(\Omega_{1}\Omega_{2}+c_{1}c_{2})},\\
\langle P^{2}\rangle &=
\frac{c_{3}(\Omega_{1}\Omega_{2}+\Omega_{2}^{2}+c_{1}^{2}+c_{1}c_{2})}
{2(c_{1}+c_{2})(\Omega_{1}\Omega_{2}+c_{1}c_{2})},\\
\frac{1}{2}\langle XP+PX\rangle &=
\frac{c_{3}(\Omega_{1}c_{1}-\Omega_{2}c_{2})}
{2(c_{1}+c_{2})(\Omega_{1}\Omega_{2}+c_{1}c_{2})},
\end{aligned}
\end{equation}
where $\Omega_{1}=\omega-2\lambda$, $\Omega_{2}=\omega+2\lambda$, $c_{1}= 6\rho+\kappa_{1}$, $c_{2}=2\rho+\kappa_{1}$, and $c_{3}=4\rho+\kappa_{1}$. The number and squeezing fluctuations are related to the covariance matrix elements as
\begin{equation}
\begin{aligned}
\delta n &= \frac{1}{2}\big(\langle X^{2}\rangle+\langle P^{2}\rangle-1\big),\\
\mathrm{Re}[\delta m] &= \frac{1}{2}\big(\langle X^{2}\rangle-\langle P^{2}\rangle\big),\\
\mathrm{Im}[\delta m] &= \frac{1}{2}\langle XP+PX\rangle.
\end{aligned}
\end{equation}
For a Gaussian bosonic state, the purity is given by~\cite{serafini2023quantum}
\begin{equation}
\mu=\frac{1}{\sqrt{4\langle X^{2}\rangle \langle P^{2}\rangle - \langle XP+PX \rangle^{2} }}.
\end{equation}
Expanding these expressions near the critical point $\lambda=\lambda_{c}$, we find results identical to those obtained from the Keldysh path-integral approach in the main text, as summarized in Table~\ref{Table_summary_1}.

\section*{APPENDIX B: One-loop calculation}
We present the one-loop correction arising from the quartic interaction generated by the two-photon loss within the Keldysh path-integral formalism~\cite{sieberer2016keldysh}. 

The quartic contribution to the Keldysh action originating from the two-photon loss reads
\begin{equation}
S^{(4)} = i\kappa_{2} \int dt\, \mathcal{V}(t),
\end{equation}
with
\begin{equation}
\mathcal{V}(t) =
a^{*}_{c}a^{*}_{q}(a^{2}_{c}+a^{2}_{q})
-(a^{*2}_{c}+a^{*2}_{q})a_{c}a_{q}
+4a^{*}_{c}a_{c}a^{*}_{q}a_{q}.
\end{equation}
The Green function in the presence of the quartic interaction, $G^{\alpha\beta}(t_{1},t_{2}) =\langle \psi_{\alpha}(t_{1})\psi^{*}_{\beta}(t_{2})\rangle$, admits the perturbative expansion
\begin{equation}
\begin{aligned}
G^{\alpha\beta}(t_{1},t_{2})
=&\, G^{\alpha\beta}_{0}(t_{1},t_{2})
+ \langle \psi_{\alpha}(t_{1})\psi^{*}_{\beta}(t_{2}) S^{(4)} \rangle_{0} \\
&-\frac{i}{2}
\langle \psi_{\alpha}(t_{1})\psi^{*}_{\beta}(t_{2}) (S^{(4)})^{2} \rangle_{0}
+ \cdots ,
\end{aligned}
\end{equation}
where $\langle \cdots \rangle_{0}$ denotes the Gaussian average defined by the quadratic Keldysh action. The last term is of order $O(\kappa_{2}^{2})$ and corresponds to a two-loop contribution. In the following, we focus on the leading one-loop correction,
\begin{equation}
\delta G^{\alpha\beta}(t_{1},t_{2})
= i\kappa_{2} \int dt\,
\langle \psi_{\alpha}(t_{1})\psi^{*}_{\beta}(t_{2}) \mathcal{V}(t) \rangle_{0}.
\end{equation}

Using Wick’s theorem and the temporal locality of $S^{(4)}$, the correction to the retarded Green function is given by
\begin{equation}
\delta G^{R}(t_{1}-t_{2})
= \int dt\, G^{R}_{0}(t_{1}-t)\, \Sigma^{R}\, G^{R}_{0}(t-t_{2}),
\end{equation}
where $\Sigma^{R}$ is the retarded self-energy matrix. The bare retarded Green function in the Nambu-Keldysh basis is
\begin{equation}
G^{R}_{0}(t) = -i
\begin{pmatrix}
\langle a_{c}(t)a^{*}_{q}(0)\rangle_{0} &
\langle a_{c}(t)a_{q}(0)\rangle_{0} \\
\langle a^{*}_{c}(t)a^{*}_{q}(0)\rangle_{0} &
\langle a^{*}_{c}(t)a_{q}(0)\rangle_{0}
\end{pmatrix}.
\end{equation}
The self-energy is determined by equal-time contractions involving one classical and one quantum field, which are obtained from the second functional derivatives of $S^{(4)}$,

\begin{equation}
\begin{aligned}
\frac{\delta^{2} S^{(4)}}{\delta a_{c} \delta a^{*}_{q}} &=
i\kappa_{2}\left(2a^{*}_{c}a_{c}-2a^{*}_{q}a_{q}+4a^{*}_{c}a_{q}\right),\\
\frac{\delta^{2} S^{(4)}}{\delta a^{*}_{c} \delta a_{q}} &=
i\kappa_{2}\left(-2a^{*}_{c}a_{c}+2a^{*}_{q}a_{q}+4a^{*}_{q}a_{c}\right),\\
\frac{\delta^{2} S^{(4)}}{\delta a_{c} \delta a_{q}} &=
i\kappa_{2}\left(-a^{*2}_{c}-a^{*2}_{q}+4a^{*}_{c}a^{*}_{q}\right),\\
\frac{\delta^{2} S^{(4)}}{\delta a^{*}_{c} \delta a^{*}_{q}} &=
i\kappa_{2}\left(a^{2}_{c}+a^{2}_{q}+4a_{c}a_{q}\right).
\end{aligned}
\end{equation}
Here, the one-loop contribution for the retarded Green function corresponds diagrammatically to the loop diagrams shown in Fig.~\ref{fig1}(c) of the main text, which represent the quartic-order corrections generated by the two-photon loss. 

From the Dyson equation
\begin{equation}
[G^{R}(\nu)]^{-1} = [G^{R}_{0}(\nu)]^{-1} - \Sigma^{R},
\end{equation}
we obtain the retarded Green function with the one-loop correction quoted in Eq.~(\ref{Eq_GR_Corr_1}) of the main text. Similarly, the Keldysh self-energy acquires the correction
\begin{equation}
\Sigma^{K} = 2i(\Gamma + 2\kappa_{2} n)\,\mathcal{I}_{2},
\end{equation}
which renormalizes the noise kernel. Using the corrected Keldysh action and the relations in Eq.~(\ref{Eq_nm_1}), we determine the number and squeezing fluctuations by setting the critical point $\lambda=\lambda_{c}$~\cite{torre2013keldysh}. Expanding the solutions for small $\kappa_{2}$ yields the scaling behavior summarized in Table~\ref{Table_summary_2} of the main text.

\begin{figure}[t!]
    \centering
    \includegraphics[width=1.0\linewidth]{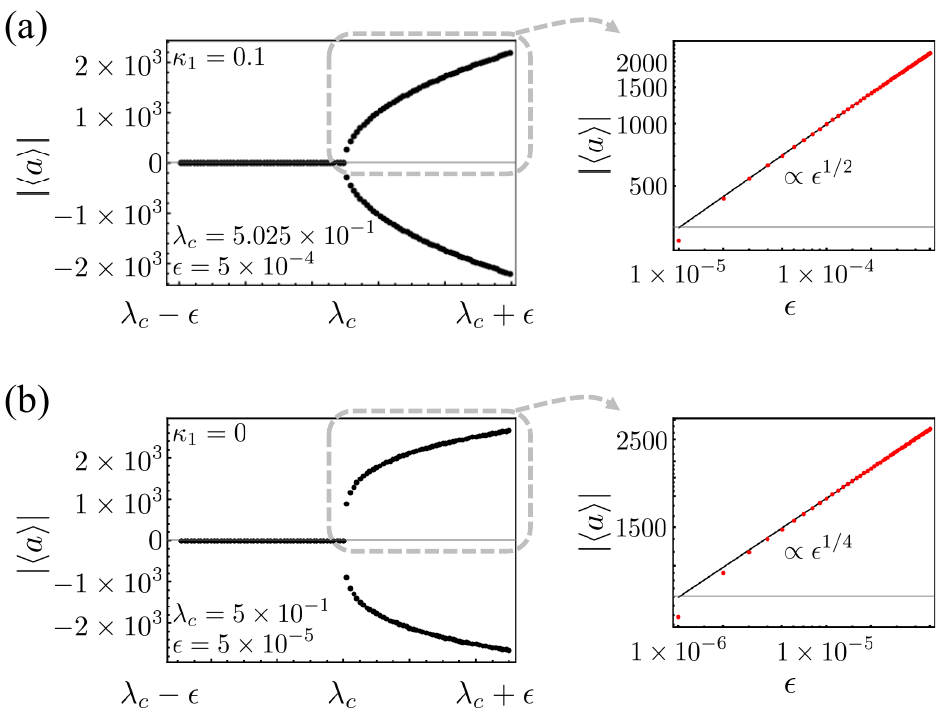}
    \caption{Numerical results for the order-parameter amplitude $|\langle a\rangle|$ as a function of the two-photon driving strength $\lambda$, obtained from the cumulant expansion. Panels (a) and (b) correspond to systems with weak parity symmetry and strong parity symmetry, respectively. The right panels show log-log plots of the superradiant phase, highlighting the scaling behavior. $\epsilon$ denotes the deviation from the critical driving strength $\lambda_{c}$. The parameters are set to $\omega = 1.0$ and $\kappa_{2} = 10^{-9}$ for all panels. }
    \label{Fig_OP}
\end{figure}

\begin{figure*}[t!]
    \centering
    \includegraphics[width=1.0\linewidth]{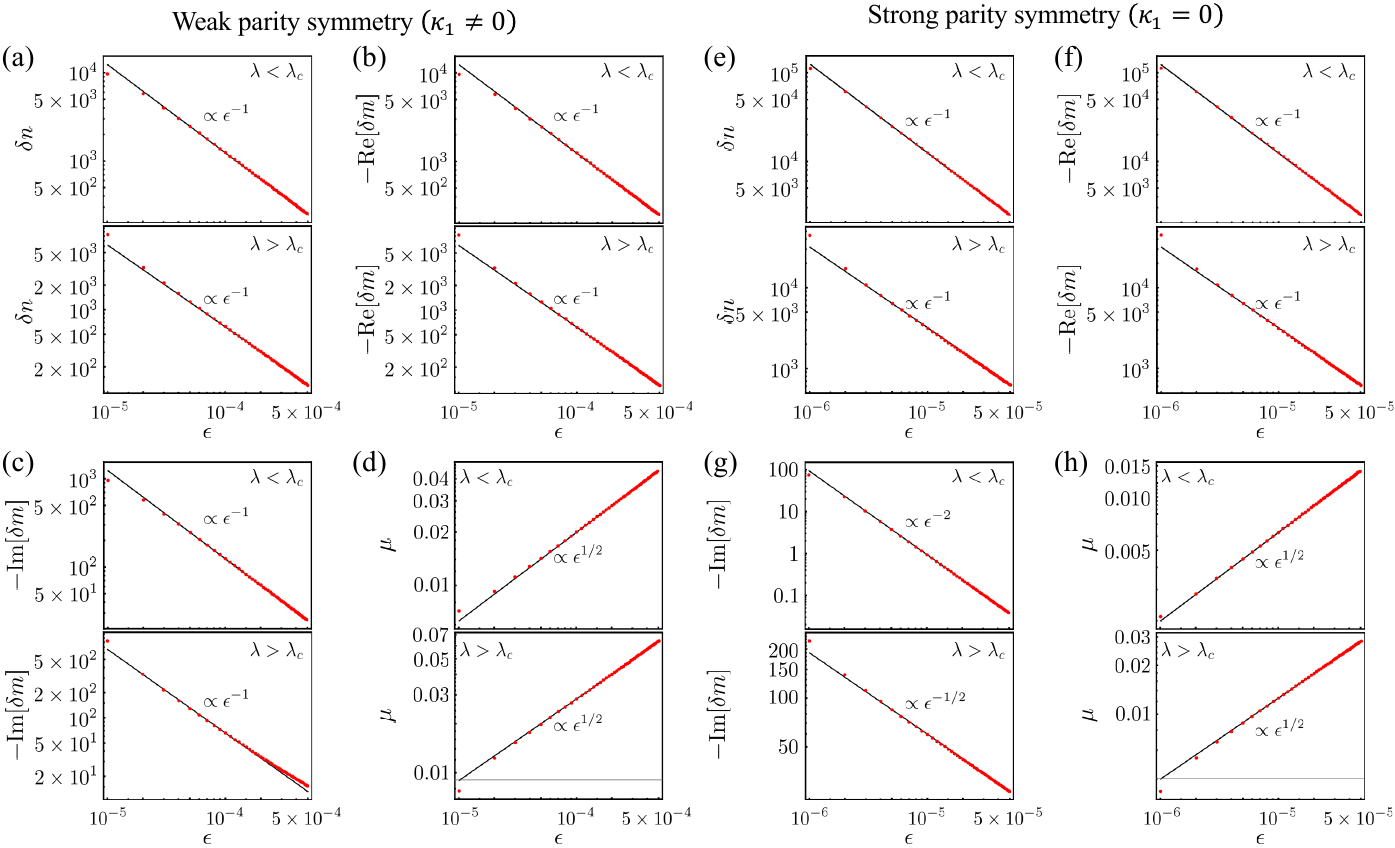}
    \caption{Numerical results for the number fluctuation $\delta n$, the squeezing fluctuations $\mathrm{Re}[\delta m]$ and $\mathrm{Im}[\delta m]$, and the purity $\mu$ as functions of the two-photon driving strength $\lambda$, obtained from the cumulant expansion. Panels (a)--(d) show $\delta n$, $\mathrm{Re}[\delta m]$, $\mathrm{Im}[\delta m]$, and $\mu$, respectively, for a system with weak parity symmetry ($\kappa_{1} \neq 0$), while panels (e)--(h) show the corresponding quantities for a system with strong parity symmetry ($\kappa_{1} = 0$). Here, $\epsilon$ denotes the deviation from the critical driving strength $\lambda_{c}$. All panels are shown on log-log scales, highlighting the critical scaling behavior near the critical point. The parameters are set to $\omega = 1.0$ and $\kappa_{2} = 10^{-9}$ for all panels, with $\kappa_{1} = 0.1$ and $\lambda_{c} = 0.502494$ in panels (a)--(d), and $\lambda_{c} = 0.5$ in panels (e)--(h).}
    \label{Fig_Weak}
\end{figure*}

\section*{APPENDIX C: Supplemental figures - numerical data from the cumulant expansion}
In this appendix, we provide additional numerical results obtained from the cumulant expansion for the order parameter, fluctuations, and purity. Figure~\ref{Fig_OP} shows the critical behavior of the order-parameter amplitude for systems with weak and strong parity symmetries, while Fig.~\ref{Fig_Weak} present the corresponding number fluctuations, squeezing fluctuations, and purity. The log-log plots highlight the critical scaling near the transition and complement the results discussed in the main text.

\section*{APPENDIX D: Detailed calculation of the cumulant expansion}
We employ a cumulant expansion to incorporate higher-order correlations and to verify the non-Gaussian critical scaling behavior. We start from the Lindblad master equation introduced in the main text as follows.
\begin{equation}
\frac{\partial \rho}{\partial t} = \mathcal{L}[\rho] = -i[H,\rho]+ \sum_{j=1,2} \mathcal{D}[L_{j}],
\end{equation}
where
\begin{equation}
    H = \omega a^{\dagger}a + \lambda(a^{2}+a^{\dagger 2}),
\end{equation}
$L_{1}=\sqrt{\kappa_{1}}a$, $L_{2}=\sqrt{\kappa_{2}}a^{2}$, and $\mathcal{D}[L]=2L\rho L^{\dagger}-L^{\dagger}L\rho - \rho L^{\dagger}L$. From this, we derive the following equations of motion for the first and second moments. 
\begin{equation}
\begin{aligned}
\frac{d\langle a\rangle}{dt} =& -(i\omega+\kappa_{1})\langle a\rangle -2i\lambda \langle a\rangle^{*} -2\kappa_{2}\langle a^{\dagger}a^{2}\rangle,\\
\frac{d m}{dt}=& -2(i\omega+\kappa_{1}+\kappa_{2})m -2i\lambda -4i\lambda n\\
&-4\kappa_{2}\langle a^{\dagger}a^{3} \rangle,\\
\frac{d n}{dt} =& +2i\lambda(m - m^{*}) -2\kappa_{1} n -4\kappa_{2}\langle a^{\dagger 2}a^{2} \rangle.
\end{aligned}
\end{equation}
To close the hierarchy, we apply a cumulant expansion and truncate the connected correlators beyond second order, which provides a controlled approximation capturing leading non-Gaussian corrections near the critical point~\cite{fowler2023determining,reiter2020cooperative}. The general formulae for three- and four-point correlators are given by
\begin{equation}
\begin{aligned}
\langle \mathcal{O}_{1}\mathcal{O}_{2}\mathcal{O}_{3} \rangle \simeq& \langle \mathcal{O}_{1}\mathcal{O}_{2}\rangle \langle \mathcal{O}_{3}\rangle + \langle \mathcal{O}_{1}\mathcal{O}_{3}\rangle \langle \mathcal{O}_{2}\rangle \\
&+ \langle \mathcal{O}_{2}\mathcal{O}_{3}\rangle \langle \mathcal{O}_{1}\rangle
-2\langle \mathcal{O}_{1}\rangle \langle \mathcal{O}_{2}\rangle \langle \mathcal{O}_{3}\rangle,
\end{aligned}
\end{equation}
and
\begin{widetext}
\begin{equation}
\begin{aligned}
\langle \mathcal{O}_{1}\mathcal{O}_{2}\mathcal{O}_{3}\mathcal{O}_{4} \rangle \simeq& \langle \mathcal{O}_{1}\mathcal{O}_{2}\rangle \langle \mathcal{O}_{3}\mathcal{O}_{4}\rangle + \langle \mathcal{O}_{1}\mathcal{O}_{3}\rangle \langle \mathcal{O}_{2}\mathcal{O}_{4}\rangle + \langle \mathcal{O}_{1}\mathcal{O}_{4}\rangle \langle \mathcal{O}_{2}\mathcal{O}_{3}\rangle
+ \langle \mathcal{O}_{1}\rangle \langle \mathcal{O}_{2}\mathcal{O}_{3}\mathcal{O}_{4}\rangle+ \langle \mathcal{O}_{2}\rangle \langle \mathcal{O}_{1}\mathcal{O}_{3}\mathcal{O}_{4}\rangle\\ 
&+ \langle \mathcal{O}_{1}\mathcal{O}_{2}\mathcal{O}_{4}\rangle \langle \mathcal{O}_{3}\rangle + \langle \mathcal{O}_{1}\mathcal{O}_{2}\mathcal{O}_{3}\rangle \langle \mathcal{O}_{4}\rangle -2 \langle \mathcal{O}_{1}\rangle \langle \mathcal{O}_{2}\rangle \langle \mathcal{O}_{3}\mathcal{O}_{4}\rangle -2 \langle \mathcal{O}_{1}\rangle \langle \mathcal{O}_{2}\mathcal{O}_{3}\rangle \langle \mathcal{O}_{4}\rangle\\
&-2 \langle \mathcal{O}_{1}\rangle \langle \mathcal{O}_{2}\mathcal{O}_{4}\rangle \langle \mathcal{O}_{3}\rangle 
-2 \langle \mathcal{O}_{1}\mathcal{O}_{2}\rangle \langle \mathcal{O}_{3}\rangle \langle \mathcal{O}_{4}\rangle 
-2 \langle \mathcal{O}_{1}\mathcal{O}_{3}\rangle \langle \mathcal{O}_{2}\rangle \langle \mathcal{O}_{4}\rangle 
-2 \langle \mathcal{O}_{1}\mathcal{O}_{4}\rangle \langle \mathcal{O}_{2}\rangle \langle \mathcal{O}_{3}\rangle\\
&+6\langle \mathcal{O}_{1}\rangle \langle \mathcal{O}_{2}\rangle \langle \mathcal{O}_{3}\rangle \langle \mathcal{O}_{4}\rangle,
\end{aligned}
\end{equation}
\end{widetext}
where $\mathcal{O}_{j=1,2,3,4}$ is the arbitrary operator. Applying these relations, the higher-order correlators appearing in the above equations of motion reduce to
\begin{equation}
\begin{aligned}
\langle a^{\dagger}a^{2}\rangle \simeq& +2n\langle a\rangle+ m \langle a\rangle^{*} -2\langle a\rangle |\langle a\rangle|^{2} ,\\
\langle a^{\dagger}a^{3}\rangle \simeq& -2\langle a\rangle|\langle a\rangle|^{2} +3n m ,\\
\langle a^{\dagger 2}a^{2}\rangle \simeq& +|m|^{2}+2n^{2}-2|\langle a\rangle|^{4} .
\end{aligned}
\end{equation}
Assuming a steady state, we obtain a closed set of coupled equations for the three variables $\langle a\rangle$, $n$, and $m$.
\begin{equation}
\begin{aligned}
0=& -(i\omega+\kappa_{1})\langle a\rangle -2i\lambda \langle a\rangle^{*} -2\kappa_{2} (+2n\langle a\rangle\\
&+ m \langle a\rangle^{*} -2\langle a\rangle |\langle a\rangle|^{2}),\\
0=& -2(i\omega+\kappa_{1}+\kappa_{2})m -2i\lambda -4i\lambda n\\
&-4\kappa_{2} (-2\langle a\rangle|\langle a\rangle|^{2} +3n m),\\
0=& +2i\lambda(m - m^{*}) -2\kappa_{1} n -4\kappa_{2} (|m|^{2}+2n^{2}\\
&-2|\langle a\rangle|^{4}).
\end{aligned}
\end{equation}
Note that these equations explicitly depend on the effective scaling parameter $\kappa_{2}$, and therefore the resulting solutions capture corrections beyond the Gaussian limit. By numerically solving these equations, we obtain the results shown in the main text and the Appendix.

We note that for any finite $\kappa_{2}$, the exact steady state of the full Lindblad dynamics preserves parity symmetry and therefore satisfies $\langle a\rangle = 0$. Accordingly, the nonzero solutions for $\langle a\rangle$ obtained from the second-order cumulant equations are not steady states for finite $\kappa_{2}$, but rather should be understood as metastable states and their fluctuation around them. Crucially, such metastable solutions become asymptotically stable in the thermodynamic limit $\kappa_{2}\to 0$, where their lifetime diverges.

\bibliography{BibRef}

\end{document}